\documentclass[a4paper,conference]{IEEEtran}
\usepackage{graphicx}
\usepackage{amsmath,amssymb}
\usepackage{booktabs}
\usepackage{microtype}
\usepackage{url}
\usepackage[hidelinks]{hyperref}
\usepackage{orcidlink}
\usepackage{balance}
\usepackage{tikz}
\usepackage[english]{babel}
\usetikzlibrary{arrows.meta,positioning,shadows.blur,fit}
\usepackage{xcolor} 

\hypersetup{
  pdftitle={Human-Certified Module Repositories for the AI Age},
  pdfauthor={Szilárd Enyedi},
  pdfkeywords={Software supply-chain security, Provenance and attestation, AI-assisted software engineering, Certified module repositories, Secure compositional architectures},
  pdfsubject={AQTR 2026},
  pdfcreator={LaTeX with hyperref},
  pdfproducer={Overleaf / TeX Live}
}

\title{Human-Certified Module Repositories for the AI Age}

\author{
    \IEEEauthorblockN{Szilárd Enyedi\,\orcidlink{0000-0002-1526-2666}}
    \IEEEauthorblockA{
        Automation Department, Technical University of Cluj-Napoca, Romania\\
    }
}

\begin{document}
\maketitle


\begingroup
\footnotesize
\noindent
\hspace{0.5em}%
\vrule width 0.6pt
\hspace{0.75em}%
\begin{minipage}{0.92\linewidth}
\textit{Preprint notice:}
This is a preprint version of a paper accepted for publication in the
Proceedings of IEEE AQTR 2026.

\vspace{0.4em}

\copyright~2026 IEEE. Personal use of this material is permitted. Permission from IEEE must be obtained for all other uses, in any current or future media, including reprinting/republishing this material for advertising or promotional purposes, creating new collective works, for resale or
redistribution to servers or lists, or reuse of any copyrighted component of this work in other works.
\end{minipage}

\vspace{0.75em}
\endgroup

\begin{abstract}
Human-Certified Module Repositories (HCMRs) are introduced in this work as a new architectural model for constructing trustworthy software in the era of AI-assisted development. As large language models increasingly participate in code generation, configuration synthesis, and multi-component integration, the reliability of AI-assembled systems will depend critically on the trustworthiness of the building blocks they use. Today's software supply-chain incidents and modular development ecosystems highlight the risks of relying on components with unclear provenance, insufficient review, or unpredictable composition behavior.

We argue that future AI-driven development workflows require repositories of reusable modules that are curated, security-reviewed, provenance-rich, and equipped with explicit interface contracts. To this end, we propose HCMRs, a framework that blends human oversight with automated analysis to certify modules and support safe, predictable assembly by both humans and AI agents. We present a reference architecture for HCMRs, outline a certification and provenance workflow, analyze threat surfaces relevant to modular ecosystems, and extract lessons from recent failures. We further discuss implications for governance, scalability, and AI accountability, positioning HCMRs as a foundational substrate for reliable and auditable AI-constructed software systems.
\end{abstract}

\begin{IEEEkeywords}
Software supply-chain security, 
Provenance and attestation, 
AI-assisted software engineering, 
Certified module repositories, 
Secure compositional architectures
\end{IEEEkeywords}

\section{Introduction}

Software development is undergoing a profound transition driven by the rapid rise of AI-assisted and AI-automated code generation. Large language models (LLMs) now routinely synthesize non-trivial software components, integrate libraries, generate configuration files, and orchestrate multi-step workflows. While this represents a breakthrough in productivity, it also amplifies long-standing risks in software engineering: opacity of generated code, silent introduction of vulnerabilities, inconsistent quality, and unpredictable integration of external dependencies.

Concurrently, the global software supply chain has entered a period of unprecedented fragility. High-impact incidents --- including the SolarWinds compromise, the Log4Shell vulnerability, and the XZ/\texttt{liblzma} backdoor --- have demonstrated that attackers increasingly target build infrastructure, update channels, and dependency graphs upon which modern development relies \cite{Willett2021SolarWindsSurvival,Martinez2021SolarWindsCaseStudy,Singh2026Log4ShellLongitudinal,Hiesgen2024Log4jTNSM}. These incidents underscore a systemic truth: today, the trustworthiness of software is governed less by local functional correctness and more by the provenance and integrity of the entire build and distribution pipeline.

In parallel, engineering practice continues to shift toward modularity, compositional automation, and reuse. Trigger--action platforms (e.g., IFTTT), flow-based environments (e.g., Node-RED), and cloud-native Infrastructure-as-Code (IaC) ecosystems (e.g., Azure Verified Modules) place composition, rather than monolithic development, at the center of software construction \cite{Mi2017IFTTTIMC,AVM_MicrosoftLearn,AVM_Portal,AVM_GitHubRepo,AVM_SolutionDev,AVM_MonthlyUpdateJan2024}. As AI tools accelerate this trend, the absence of curated, provenance-rich, and security-reviewed building blocks becomes a first-order risk.

\textbf{This paper introduces Human-Certified Module Repositories (HCMRs)} as a new architectural model for safe, verifiable, and auditable AI-constructed software systems. An HCMR is a curated repository of reusable, parameterized, and security-reviewed components that provide strong provenance, explicit interface contracts, and---where appropriate---formal guarantees. By constraining both humans and AI agents to assemble software from certified modules, HCMRs aim to deliver predictable behavior, compositional safety, and end-to-end auditability.

\textbf{Contributions.} Rather than merely surveying prior work, this paper proposes HCMRs as a concrete foundation for secure composition:
\begin{enumerate}
    \item \textbf{Concept and motivation:} We define HCMRs and motivate their necessity at the intersection of AI-assisted development and supply-chain risk, synthesizing lessons from recent incidents and modular ecosystems.
    \item \textbf{Reference architecture:} We present a design for HCMRs, including certification workflow, assurance tiers, machine-readable metadata, and contract-aware composition constraints suitable for IDE and agentic tooling.
    \item \textbf{Threat model and mitigations:} We articulate adversary capabilities and attack surfaces relevant to module ecosystems and show how HCMRs address them via provenance, human certification, and safe assembly rules.
    \item \textbf{Grounding and implications:} We relate HCMRs to adjacent paradigms (formal verification, SLSA-style provenance, identity-based signing) and discuss implications for governance, scalability, and adoption.
    \item \textbf{Open problems:} We outline research challenges in compositional safety, scalable certification, developer usability, and AI accountability.
\end{enumerate}

The remainder of the paper develops this proposal in depth: we review background and related work to position HCMRs, formalize the threat model, detail the architecture and certification pipeline, analyze representative case studies, and conclude with a discussion of limitations and open research directions.

The central insight is straightforward: future software will be built by AI, and we must ensure that the building blocks it uses are trustworthy. HCMRs provide the human certification needed to make this possible.
\section{Background and Motivation}

Modern software systems increasingly depend on deep dependency graphs, distributed build pipelines, and complex integration surfaces. This creates a structural problem: the trustworthiness of software now depends less on local code correctness and more on the integrity of the entire supply chain that produces and distributes it. The SolarWinds compromise illustrated this vividly, when attackers injected malicious code into Orion platform updates downloaded by roughly 18\,000 downstream organizations, including multiple U.S. federal agencies. Academic analyses highlight that the breach was enabled by weaknesses in update-signing trust paths and insufficient isolation in build systems, showing how state‑level actors can weaponize supply‑chain interdependencies at scale \cite{Coco2022SolarWindsEJIL,Ghanbari2024SolarWindsKaseyaTeachingCase}.

Another major turning point was the Log4Shell vulnerability (CVE‑2021‑44228), which was rapidly exploited worldwide due to the ubiquity of the Log4j logging library. Long‑term telescope‑based measurements show that exploitation persisted for years, gradually consolidating around stable attacker infrastructures while scanning behavior evolved in sophistication \cite{Singh2026Log4ShellLongitudinal}. A complementary network‑forensic study in IEEE Transactions confirmed that malicious scanning continued in waves throughout 2022 and beyond, long after initial patching efforts subsided \cite{Hiesgen2024Log4jTNSM}. The key insight is that once a widely reused component becomes vulnerable, ecosystem‑level exposure persists long after remediation.

Most recently, the 2024–2025 XZ Utils backdoor (CVE‑2024‑3094) demonstrated that supply‑chain attacks can originate not only from compromised CI pipelines but from social engineering campaigns targeting volunteer maintainers. An arXiv analysis documented the multi‑stage infection path: subtle changes to build scripts, hidden payloads embedded in test files, and an activation mechanism targeting \texttt{sshd}, enabling remote code execution without authentication \cite{Lins2024XZArxiv}. A 2025 LNCS paper further showed that \texttt{xz} is a dependency for nearly 30\,000 Debian and Ubuntu packages, making this compromise comparable in ecosystem importance to \texttt{glibc} and demonstrating the catastrophic blast radius of a single compromised maintainer account \cite{Lins2025XZLNCS}.

Parallel to these attacks, formal verification research has advanced the state of high‑assurance software. The seL4 microkernel remains the most prominent example: a general‑purpose OS kernel with a machine‑checked proof of functional correctness from C source to binary, maintained over nearly a decade of evolution \cite{Klein2014seL4TOCS}. Earlier SOSP work established seL4 as the first feasibility demonstration of a fully verified kernel with performance comparable to high‑performance L4 family kernels \cite{Klein2009seL4SOSP}. These results demonstrate that for small and security‑critical components, mathematical verification is achievable — but also highly resource‑intensive.

Similarly, the CompCert C compiler showed that verified compilation is possible for real‑world subsets of C. Leroy’s CACM article demonstrated a mechanized proof of semantic preservation from Clight to PowerPC assembly, eliminating entire classes of miscompilation bugs \cite{Leroy2009CompCertCACM}. CompCert’s guarantees provide strong assurance for safety‑critical systems where compiler correctness is part of the trust base.

While formal verification is effective for small components, it does not scale to entire application ecosystems. A complementary direction has emerged through provenance and signing infrastructures—particularly SLSA (Supply‑chain Levels for Software Artifacts). The SLSA provenance specification defines structured attestation formats that bind software artifacts to their build process, including builder identity, build instructions, environment parameters, and dependency digests \cite{SLSAProvenanceV01,SLSAProvenanceV02}. Recent SLSA guides emphasize end‑to‑end integrity, recommending automated provenance generation and verification as core defenses against supply‑chain manipulation \cite{PracticalDevSecOps2026SLSAGuide,SLSAGitHubGenerator}.

In the signing domain, Sigstore introduced an identity‑based signing model using ephemeral keys authenticated via OIDC, together with a transparency log (Rekor) that provides globally auditable signatures. Its CCS’22 paper argues that this drastically reduces key‑management burdens while improving artifact auditability across ecosystems \cite{Newman2022SigstoreCCS,SigstoreOverviewDocs}. Usability research has further shown that tool adoption is shaped by workflow integration challenges, highlighting developer experience as a central factor in secure‑by‑default signing ecosystems \cite{Kalu2025SigstoreUsability}.

Finally, software production is increasingly modular \cite{AVM_MicrosoftLearn,AVM_Portal,AVM_GitHubRepo,AVM_SolutionDev,AVM_MonthlyUpdateJan2024}. Trigger‑action platforms like IFTTT have grown rapidly, enabling end‑users to compose automation workflows from thousands of reusable components. An IMC’17 measurement study quantified the ecosystem’s expansion, showing that more than half of all available services relate to IoT devices, with pronounced inefficiencies in trigger latency and execution pathways \cite{Mi2017IFTTTIMC,MiIFTTTdatasetGitHub}. In parallel, flow‑based programming tools such as Node‑RED provide visual composition primitives widely used in IoT systems; empirical studies demonstrate their value for rapid integration but highlight weaknesses in systematic testing and formal assurance \cite{IEEE2018NodeREDHomeAutomation}.

These trajectories — escalating supply‑chain threats, maturing provenance frameworks, breakthroughs in verification, and explosive growth in composable software modules — jointly motivate the emergence of Human‑Certified Module Repositories (HCMRs). The next section situates HCMRs within the broader landscape of related work.
\section{Related Foundations and Ecosystems}

This section surveys prior work in five domains that form the intellectual foundation for Human‑Certified Module Repositories (HCMRs): formal verification ecosystems, supply‑chain attack analyses, provenance and signing infrastructures, trigger‑action ecosystems, and cloud‑native module libraries.

\subsection{Formal Verification Ecosystems}

The seL4 microkernel remains the most significant achievement in practical formal verification. The TOCS 2014 article provides a comprehensive treatment of kernel‑level proofs, including binary‑level correctness, information‑flow noninterference, and verified access‑control enforcement, all backed by machine‑checked proofs in Isabelle/HOL \cite{Klein2014seL4TOCS}. Earlier work presented at SOSP’09 established seL4 as the first general‑purpose kernel with a full functional correctness proof from C to specification, demonstrating feasibility at roughly 10\,000 lines of C code \cite{Klein2009seL4SOSP}. These works show that formal verification can eliminate entire classes of implementation bugs in small but highly critical components.

The CompCert verified C compiler provides another major pillar. Leroy’s CACM paper introduced a machine‑checked proof of semantic preservation for a realistic subset of C and demonstrated that verified compilation can outperform mainstream compilers in predictability and correctness \cite{Leroy2009CompCertCACM}. The material illustrates the methodological framework behind verified compilation, showing how correctness proofs can be decomposed into trace‑preservation lemmas across compilation passes.

While seL4 and CompCert show that high‑assurance components are achievable, the resource cost is substantial. These insights motivate hybrid approaches that combine strong curation (as proposed for HCMRs) with selective verification of module contracts rather than entire applications.

\subsection{Supply‑Chain Attack Literature}

Academic work on the SolarWinds breach provides an empirical foundation for understanding modern supply‑chain compromises. Willett’s analysis in \emph{Survival} documents how attackers inserted malicious code into SolarWinds Orion updates and leveraged signed distribution channels to propagate the backdoor to thousands of organizations, illustrating the systemic risk posed by trusted update infrastructures \cite{Willett2021SolarWindsSurvival}. A complementary study in the \emph{International Journal of Safety and Security Engineering} highlights architectural issues in third‑party dependencies and recommends SBOMs, zero‑trust architectures, and MFA as practical mitigations \cite{Martinez2021SolarWindsCaseStudy}.

Log4Shell analyses provide further context on vulnerability propagation \cite{Pauley2023CVEWaybackIMC}. The 2026 arXiv longitudinal study shows multi‑year exploitation patterns, infrastructure consolidation, and persistent attacker activity targeting vulnerable systems, underscoring the difficulty of mitigating ecosystem‑wide exposure once a vulnerability becomes widely weaponized \cite{Singh2026Log4ShellLongitudinal}. The 2024 IEEE Transactions study adds detailed scanning behavior, demonstrating repeated scanning waves and dominance of specific malicious actors in sustained exploitation campaigns \cite{Hiesgen2024Log4jTNSM}.

The recent XZ Utils backdoor expands the literature by revealing a highly sophisticated supply‑chain compromise introduced through targeted social engineering of maintainers. The arXiv study provides an attack‑path analysis showing how minor build‑script modifications enabled injection of hidden malicious payloads \cite{Lins2024XZArxiv}. The LNCS 2025 paper emphasizes the ecosystem‑level blast radius by quantifying how \texttt{xz} is embedded in tens of thousands of downstream packages in major Linux distributions, highlighting the fragility of volunteer‑maintained critical components \cite{Lins2025XZLNCS}.

Collectively, these analyses demonstrate that supply‑chain attacks require comprehensive mitigations across build systems, trust chains, maintainership models, and dependency governance — all core concerns for HCMRs.

\subsection{Provenance and Signing Frameworks}

SLSA (Supply‑chain Levels for Software Artifacts) defines a maturity model and attestation framework for verifying software provenance. The official specification describes provenance predicates capturing builder identity, build instructions, parameters, environment variables, and dependency digests, structured using in‑toto attestations and DSSE envelopes \cite{SLSAProvenanceV01,SLSAProvenanceV02}. Recent practitioner guides stress automated, tamper‑resistant build services and cryptographically verifiable provenance as foundations for modern supply‑chain security, motivated by large‑scale incidents such as SolarWinds and Log4Shell \cite{PracticalDevSecOps2026SLSAGuide,SLSAGitHubGenerator}.

Sigstore represents a complementary approach focused on developer‑friendly artifact signing. The CCS’22 paper introduces ephemeral key signing tied to OIDC identities and verified through the Rekor transparency log, significantly reducing key‑management burdens and improving auditability across open‑source ecosystems \cite{Newman2022SigstoreCCS,SigstoreOverviewDocs}. Usability studies show that developers' adoption decisions depend strongly on workflow integration and tooling ergonomics, indicating that secure signing must also be easy to deploy and maintain \cite{Kalu2025SigstoreUsability}.

These frameworks strongly influence the HCMR design: module certification requires trustworthy provenance and cryptographically authenticated distribution to ensure integrity and traceability.

\subsection{Trusted Data Repositories}

A related but fundamentally different lineage of work concerns the certification of trusted repositories for digital data. Frameworks such as CoreTrustSeal \cite{CoreTrustSeal2024ICPSR} define organizational and technical requirements for trustworthy data stewardship, including governance practices, sustainability models, metadata quality, and long-term preservation guarantees. Similarly, the ISO~16363 standard \cite{ISO16363_2025} provides a formal audit methodology for assessing digital preservation infrastructures, emphasizing durability, institutional policies, and conformance with archival reference models.

While these initiatives establish important foundations for repository trustworthiness, they are designed for static digital assets rather than software modules. They do not certify code, behavioral interfaces, dependency graphs, or build provenance, and they provide no mechanisms for ensuring secure or correct composition of software components. In contrast, Human-Certified Module Repositories (HCMRs) address the dynamic, operational, and security-sensitive nature of software modules and are motivated by challenges such as supply-chain compromise, AI-driven assembly, and the need for verifiable build processes. As such, HCMRs extend the idea of “trusted repositories” beyond archival data preservation into the domain of actively curated, provenance-aware, composable software ecosystems.

\subsection{Trigger–Action and Flow‑Based Programming Ecosystems}

IFTTT has been extensively studied as a large‑scale trigger–action ecosystem. An IMC’17 measurement study documented rapid ecosystem expansion, with IoT‑related services comprising more than half of all available channels, and revealed significant latency and execution-path inefficiencies due to polling‑based triggers and cloud intermediation \cite{Mi2017IFTTTIMC,MiIFTTTdatasetGitHub}. These insights highlight both the promise and the limitations of lightweight, end‑user integrator models.

Flow‑based programming environments like Node‑RED further decentralize system assembly. A 2018 ACM paper proposed systematic testing and UML‑based modeling for Node‑RED flows, noting that the inherent flexibility of visual wiring can reduce understandability and obscure error propagation paths in IoT applications \cite{Clerissi2018NodeRedTesting}. Additional IoT‑analytics papers show Node‑RED’s value for rapid experimentation but reinforce the need for structured development methodologies in safety‑critical domains \cite{Onwuegbuzie2024NodeRedAnalytics,IEEE2018NodeREDHomeAutomation}.

These ecosystem studies demonstrate that modular composition frameworks succeed when modules are discoverable, interoperable, and easy to integrate — but also that lack of verification and curation leads to fragility. HCMRs aim to preserve the composability benefits while adding structured certification.

\subsection{Cloud‑Native Module Libraries}

Azure Verified Modules (AVM) exemplify a modern, curated module ecosystem. Microsoft Learn documentation describes AVM as a unified standard for well‑architected IaC modules with strict coding guidelines, automated testing, versioning, and consistent interfaces across Bicep and Terraform \cite{AVM_MicrosoftLearn}. The official AVM project site further emphasizes cross‑language alignment, composability, and role as a trusted foundation for enterprise‑grade deployments, supported directly by Microsoft engineering teams \cite{AVM_Portal,AVM_GitHubRepo,AVM_SolutionDev,AVM_MonthlyUpdateJan2024}.

AVM demonstrates that large‑scale curated module ecosystems can enforce consistent design and reliability, serving as an industrial precedent for the more security‑focused HCMR model.

Table~\ref{tab:ecosystems} compares governance, provenance, and certification across modular ecosystems. 

\begin{table}[!t]
\centering
\caption{Comparison of governance, provenance, and certification properties across modular ecosystems.}
\label{tab:ecosystems}
\begin{tabular}{lccc}
\toprule
\textbf{Ecosystem} & \textbf{Governance} & \textbf{Provenance} & \textbf{Certification} \\
\midrule
IFTTT           & Low    & None    & None   \\
Node-RED        & Medium & None    & None   \\
AVM (Azure)     & High   & Partial & Strong \\
HCMR (Proposed) & High   & Strong  & Strong \\
\bottomrule
\end{tabular}   
\end{table}

\section{Problem Statement and Core Requirements}

Modern software systems exhibit increasing structural fragility due to the combination of deep dependency trees, heterogeneous build infrastructures, and a globalized open-source ecosystem. High-profile supply-chain compromises such as SolarWinds demonstrated that a single compromised build system can propagate malicious updates to more than 18{,}000 downstream organizations, leveraging trusted digital signatures and automated update pipelines to spread the attack silently and efficiently \cite{Willett2021SolarWindsSurvival,Martinez2021SolarWindsCaseStudy}. Subsequent academic analyses have emphasized the systemic nature of the breach, showing how tampering in upstream build processes can bypass traditional endpoint security controls and undermine the security posture of thousands of organizations simultaneously \cite{Coco2022SolarWindsEJIL,Ghanbari2024SolarWindsKaseyaTeachingCase}.

The Log4Shell vulnerability (CVE-2021-44228) further revealed the long-term ecosystem exposure resulting from a widely used but insufficiently governed component. Measurements from an active network telescope show sustained exploitation activity continuing for years after disclosure, with attackers reusing infrastructure, evolving scanning approaches, and repeatedly probing unpatched systems across global vantage points \cite{Singh2026Log4ShellLongitudinal}. Additional studies in IEEE Transactions highlight that malicious scanning waves persisted throughout 2022, long after the initial disclosure period, thus demonstrating that ungoverned dependency ecosystems cannot rely solely on patch release cycles for risk mitigation \cite{Hiesgen2024Log4jTNSM}.

More recently, the XZ Utils backdoor (CVE-2024-3094) exposed the vulnerabilities inherent in volunteer-maintained critical infrastructure. Academic analyses show that a single socially engineered contributor was able to gradually gain trust and commit rights, ultimately introducing a multi-stage backdoor targeting \texttt{sshd} on thousands of Linux systems through subtle build-script manipulations and embedded payloads \cite{Lins2024XZArxiv,Kaspersky2024XZAnalysis}. A 2025 LNCS study quantified the blast radius by identifying nearly 30{,}000 Debian and Ubuntu packages that depend—directly or transitively—on the affected \texttt{xz} libraries, underscoring how fragility at the maintainer level can affect entire operating-system distributions \cite{Lins2025XZLNCS,MITSTAMP2025XZDeck}.

Despite advances in formal verification, the ecosystem remains largely unprotected. While systems like the seL4 microkernel demonstrate that full verification of critical components is possible and maintainable over long periods of evolution \cite{Klein2014seL4TOCS}, such efforts are economically infeasible for the vast majority of modules that compose modern software. Similarly, verified compilers like CompCert provide correctness guarantees for translation from C to assembly \cite{Leroy2009CompCertCACM}, yet do not address the integrity of the codebases or dependencies themselves.

Provenance standards and signing infrastructures provide partial mitigations. The SLSA provenance specification offers structured, verifiable metadata linking artifacts to builder identities, build parameters, and dependency digests \cite{SLSAProvenanceV01,SLSAProvenanceV02}. Sigstore further reduces the complexity of managing signing keys through identity-based ephemeral key signing and transparency logs, improving auditability across ecosystems \cite{Newman2022SigstoreCCS,SigstoreOverviewDocs}. However, these systems remain optional, and adoption varies widely across open-source communities.

In parallel, modular development ecosystems continue to expand rapidly. Trigger–action platforms like IFTTT now integrate hundreds of services and have seen significant growth in IoT-focused automations, but empirical studies highlight performance inefficiencies and limited governance of shared components \cite{Mi2017IFTTTIMC}. Flow-based programming frameworks such as Node-RED are widely used in IoT and automation scenarios, yet academic studies show weaknesses in testing, validation, and lifecycle assurance of flows constructed from community-contributed nodes \cite{Clerissi2018NodeRedTesting,Onwuegbuzie2024NodeRedAnalytics,IEEE2018NodeREDHomeAutomation}.

At the same time, cloud providers demonstrate that curated, high-quality module ecosystems are feasible. Azure Verified Modules (AVM), for example, enforce strict development guidelines, versioning practices, and testing standards across dozens of Infrastructure-as-Code modules maintained by Microsoft engineering teams, ensuring consistency and alignment with well-architected frameworks \cite{AVM_MicrosoftLearn,AVM_Portal,AVM_GitHubRepo,AVM_SolutionDev,AVM_MonthlyUpdateJan2024}.

Taken together, these findings reveal a core problem: the modern software ecosystem is highly modular, highly automated, and increasingly AI-assisted — but lacks a trustworthy substrate of vetted, provenance-rich, and security-reviewed modules for safe composition. Human-Certified Module Repositories (HCMRs) are proposed to fill this gap by establishing a curated, security-reviewed, and provenance-verified ecosystem of reusable modules suitable for both human and AI-driven assembly.
\section{Human-Certified Module Repositories: Architecture and Workflow}

Human-Certified Module Repositories (HCMRs) aim to provide a curated foundation for constructing software systems from trusted, reusable components. Whereas traditional package repositories prioritize scale and openness, HCMRs emphasize correctness, provenance, operational sanity, and compositional safety. Inspired by developments in formal verification, software supply-chain security, and cloud-native module ecosystems, HCMRs propose a middle ground between fully verified systems and ungoverned community repositories.

\subsection{Design Principles}

\begin{figure}[t]
\centering

\definecolor{PastelOrange}{RGB}{255,201,147} 
\definecolor{PastelRed}   {RGB}{255,179,186} 
\definecolor{PastelBlue}  {RGB}{179,205,227} 
\definecolor{PastelGreen} {RGB}{204,235,197} 
\definecolor{PastelLilac} {RGB}{222,203,228} 

\begin{tikzpicture}[
    node distance=1.1cm,
    stageBase/.style={
        rectangle, rounded corners, draw=black, thick,
        minimum height=1.15cm,
        text width=5.2cm,  
        align=center, inner xsep=6pt, inner ysep=6pt
    },
    arrow/.style={-Latex, very thick}
]

\ifdefined\slsamono
  \def\cSrc{white}\def\cBld{white}\def\cProv{white}\def\cArt{white}\def\cVer{white}
\else
  \def\cSrc{PastelOrange}\def\cBld{PastelRed}\def\cProv{PastelBlue}\def\cArt{PastelGreen}\def\cVer{PastelLilac}
\fi

\node[stageBase, fill=\cSrc]  (source)    {Source Code};
\node[stageBase, fill=\cBld,  below=of source]  (builder)  {Isolated Builder};
\node[stageBase, fill=\cProv, below=of builder] (prov)     {Provenance Generation};
\node[stageBase, fill=\cArt,  below=of prov]    (artifact) {Artifact + Attestation};
\node[stageBase, fill=\cVer,  below=of artifact] (consumer) {Verifier / Consumer};

\draw[arrow] (source.south)   -- (builder.north);
\draw[arrow] (builder.south)  -- (prov.north);
\draw[arrow] (prov.south)     -- (artifact.north);
\draw[arrow] (artifact.south) -- (consumer.north);

\end{tikzpicture}
\caption{SLSA-inspired provenance generation and attestation pipeline.}
\label{fig:slsa-pipeline}
\end{figure}
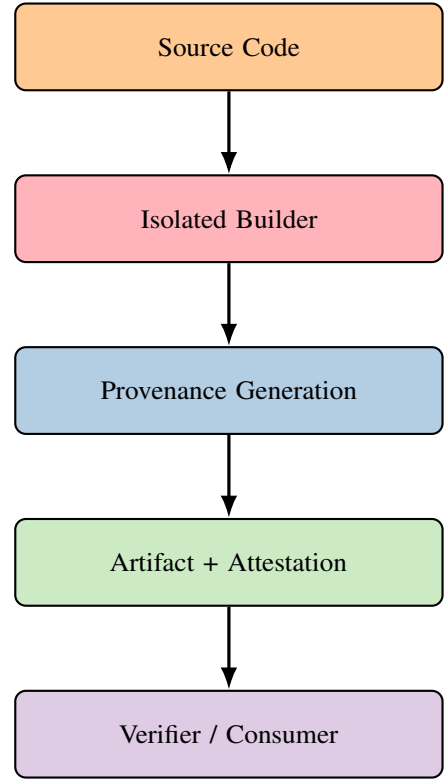

HCMRs rest on six guiding principles:

\textbf{1) Strong provenance and auditable trust chains.}  
Each module must include verifiable provenance metadata describing its origin, builder identity, build process, and dependency digests. The SLSA provenance model provides a reference template for such metadata, defining attestation structures that capture builder IDs, build configurations, and material digests through in-toto statements and DSSE envelopes \cite{SLSAProvenanceV01}. The SLSA provenance pipeline (Fig.~\ref{fig:slsa-pipeline}) illustrates the role of isolated builders.

\textbf{2) Human certification.}  
Certification teams perform security reviews, interface consistency checks, abuse-resistance analyses, and—where possible—machine-assisted formal reasoning over module behavior. This approach mirrors the strict review and approval processes in curated ecosystems such as Azure Verified Modules, where modules are systematically tested, versioned, and validated against architectural guidelines before release.

\textbf{3) Composable interfaces with explicit contracts.}  
Each module exposes well-defined inputs, outputs, and invariants. Interfaces must be machine-readable, enabling both static analysis and AI-based composition. This is particularly important given empirical findings from Node-RED and IFTTT ecosystems showing that loosely specified interfaces frequently result in orchestration failures and unexpected behavior in trigger–action workflows \cite{Mi2017IFTTTIMC}.

\textbf{4) Secure-by-default assembly constraints.}  
Composition engines—whether IDE tools or AI agents—should only be able to assemble modules that satisfy compatibility constraints, provenance checks, and dependency integrity rules. This requirement is informed by supply-chain incidents such as the SolarWinds breach, where compromised update channels propagated malicious components downstream through otherwise legitimate dependency relationships.

\textbf{5) Multi-tier assurance levels.}  
Inspired by SLSA levels and the graduated assurance models used in AVM modules, HCMRs define multiple certification levels. Lower tiers emphasize robust engineering practices and provenance; higher tiers incorporate formal reasoning, static analysis, or semi-formal specification checks \cite{PracticalDevSecOps2026SLSAGuide}.

\textbf{6) Ecosystem-scale maintainability.}  
The XZ Utils incident demonstrated the risk posed by maintainers operating under long-term social-engineering pressure, ultimately introducing subtle backdoors into critical compression libraries used by tens of thousands of downstream packages. HCMRs incorporate strong governance models to avoid single-maintainer bottlenecks.

These principles provide a structural foundation for the HCMR model, integrating supply-chain security insights, modern provenance practices, and curated ecosystem design.

\subsection{Certification Pipeline}

The certification pipeline ensures that modules entering the repository meet rigorous standards for quality, security, and composability.

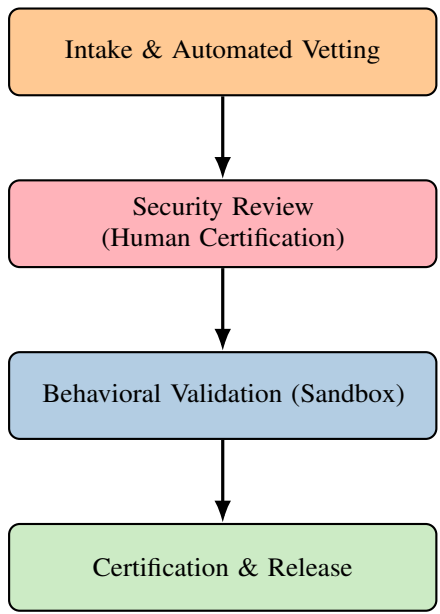
\begin{figure}[t]
\centering

\definecolor{PastelOrange}{RGB}{255,201,147} 
\definecolor{PastelRed}   {RGB}{255,179,186} 
\definecolor{PastelBlue}  {RGB}{179,205,227} 
\definecolor{PastelGreen} {RGB}{204,235,197} 

\begin{tikzpicture}[
  node distance=1.1cm, 
  stageBase/.style={
    rectangle, rounded corners, draw=black, thick,
    minimum height=1.15cm,
    text width=5.2cm, 
    align=center, inner xsep=6pt, inner ysep=6pt
  },
  arrow/.style={-Latex, very thick}
]

\ifdefined\hcmrmono
  \def\fillIntake{white}\def\fillReview{white}\def\fillValid{white}\def\fillRelease{white}
\else
  \def\fillIntake{PastelOrange}\def\fillReview{PastelRed}\def\fillValid{PastelBlue}\def\fillRelease{PastelGreen}
\fi

\node[stageBase, fill=\fillIntake]   (intake)    {Intake \& Automated Vetting};
\node[stageBase, fill=\fillReview,   below=of intake]     (review)    {Security Review (Human Certification)};
\node[stageBase, fill=\fillValid,    below=of review]     (validation){Behavioral Validation (Sandbox)};
\node[stageBase, fill=\fillRelease,  below=of validation] (release)   {Certification \& Release};

\draw[arrow] (intake.south)    -- (review.north);
\draw[arrow] (review.south)    -- (validation.north);
\draw[arrow] (validation.south) -- (release.north);

\end{tikzpicture}
\caption{HCMR certification pipeline, showing the four stages of intake, security review, behavioral validation, and final certification.}
\label{fig:hcmr-workflow}
\end{figure}

As shown in Fig.~\ref{fig:hcmr-workflow}, it consists of four stages:

\textbf{1) Intake and vetting.}  
Modules submitted to the repository undergo automated checks for dependency hygiene, reproducible builds, provenance completeness, and alignment with defined interface contracts. SLSA-aligned provenance verification is performed at this stage, ensuring linkability between source code, builder identity, and resulting artifacts \cite{SLSAProvenanceV01}.

\textbf{2) Security review.}  
Human auditors evaluate static analysis reports, inspect sensitive code paths, examine dependency graphs, and assess potential abuse cases. Supply-chain attack literature provides specific patterns to watch for, such as hidden dependency insertion, build-script tampering, and staged activation logic—behaviors that were central to the XZ Utils backdoor and SolarWinds compromise.

\textbf{3) Behavioral validation.}  
Modules are executed in sandboxed environments to validate runtime behavior across representative configurations. Techniques inspired by IoT workflow testing (e.g., Node-RED model-based validation) help detect deviations from expected behaviors, ensuring that modules behave predictably under composition.

\textbf{4) Certification and release.}  
Once validated, modules are assigned an assurance level and published together with machine-readable metadata describing their interfaces, invariants, provenance, and dependency constraints. This mirrors AVM's structured publishing model, which uses standardized repositories and automated documentation generation to ensure consistency across releases.

\subsection{Metadata Model}

HCMR metadata unifies concepts from provenance systems, interface description languages, and module certification frameworks. An HCMR metadata record includes:

\begin{itemize}
  \item \textbf{Provenance attestation:} A SLSA‑aligned provenance statement capturing builder identity, build process, and dependency digests \cite{SLSAProvenanceV01}.
  \item \textbf{Interface contract:} A structured definition of input parameters, output parameters, valid ranges, and invariant conditions.
  \item \textbf{Security attributes:} Information about required permissions, threat-model assumptions, and relevant abuse-resistant constraints derived from supply-chain case studies.
  \item \textbf{Assurance level:} A tiered classification inspired by industry frameworks (e.g., SLSA, AVM), indicating the rigor of review, testing, and formal analysis applied to the module \cite{PracticalDevSecOps2026SLSAGuide}.
  \item \textbf{Dependency graph digest:} A cryptographic summary of transitive dependencies, essential given widespread risks demonstrated in incidents like Log4Shell, where transitive dependencies created unexpected systemic exposure \cite{Singh2026Log4ShellLongitudinal}.
\end{itemize}

This metadata enables both human reviewers and automated assembly tools to reason about module integrity, compatibility, and safety.

\subsection{AI-Assisted Assembly Pipeline}

Given the rapid rise of LLM‑based developer assistants, HCMRs integrate an AI-aware assembly pipeline designed to guide automated agents toward safe compositions.

The pipeline consists of:

\textbf{1) Constraint-based module discovery.}  
AI agents are restricted to modules in the HCMR catalog and must satisfy trust-chain requirements before selecting components. This approach prevents the inadvertent ingestion of unvetted dependencies—an issue highlighted in IFTTT studies showing that open-ended composition can lead to unpredictable behavior in cross-service automations \cite{Mi2017IFTTTIMC}.

\textbf{2) Contract-based synthesis.}  
Model prompts include interface contracts and invariants extracted from the metadata model. This ensures that generated workflows satisfy type, range, and dependency constraints. The formal verification insights from seL4 demonstrate the importance of clearly defined and enforced invariants in ensuring predictable system behavior \cite{Klein2014seL4TOCS}.

\textbf{3) Provenance-aware build orchestration.}  
The build system automatically generates and verifies SLSA-aligned provenance for all composed artifacts. Provenance requirements ensure that final outputs can be traced back to certified modules and verifiable build paths, reflecting best practices recommended in the SLSA framework and associated guidelines \cite{PracticalDevSecOps2026SLSAGuide}.

\begin{figure}[t]
\centering

\definecolor{PastelOrange}{RGB}{255,201,147} 
\definecolor{PastelRed}   {RGB}{255,179,186} 
\definecolor{PastelBlue}  {RGB}{179,205,227} 
\definecolor{PastelGreen} {RGB}{204,235,197} 
\definecolor{PastelLilac} {RGB}{222,203,228} 

\begin{tikzpicture}[
    node distance=1.1cm,
    boxBase/.style={
        rectangle, rounded corners, draw=black, thick,
        minimum height=1.15cm,
        text width=5.2cm,  
        align=center, inner xsep=6pt, inner ysep=6pt
    },
    arrow/.style={-Latex, very thick}
]

\ifdefined\sigmono
  \def\cOIDC{white}\def\cFulcio{white}\def\cSign{white}\def\cRekor{white}\def\cVerify{white}
\else
  \def\cOIDC{PastelOrange}\def\cFulcio{PastelRed}\def\cSign{PastelBlue}\def\cRekor{PastelGreen}\def\cVerify{PastelLilac}
\fi

\node[boxBase, fill=\cOIDC]   (oidc)   {OIDC Identity Provider};
\node[boxBase, fill=\cFulcio, below=of oidc]  (fulcio) {Fulcio (Ephemeral Certificate)};
\node[boxBase, fill=\cSign,   below=of fulcio] (sign)   {Artifact Signing};
\node[boxBase, fill=\cRekor,  below=of sign]   (rekor)  {Rekor Transparency Log};
\node[boxBase, fill=\cVerify, below=of rekor]  (verify) {Verification};

\draw[arrow] (oidc.south)   -- (fulcio.north);
\draw[arrow] (fulcio.south) -- (sign.north);
\draw[arrow] (sign.south)   -- (rekor.north);
\draw[arrow] (rekor.south)  -- (verify.north);

\end{tikzpicture}
\caption{Identity-based signing and transparency logging workflow inspired by Sigstore.}
\label{fig:sigstore-flow}
\end{figure}
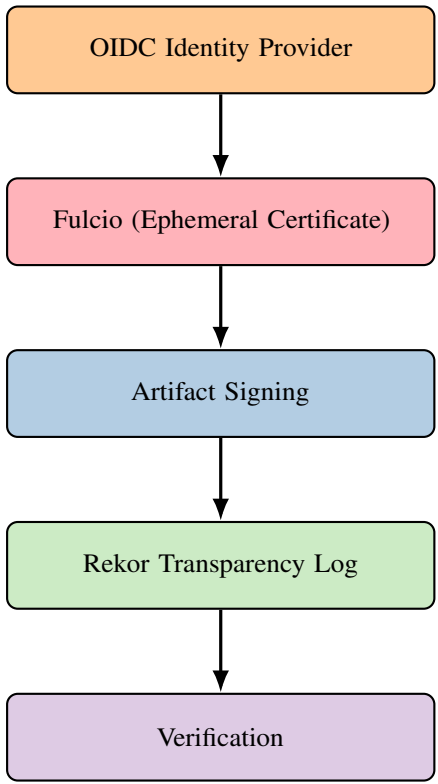

\textbf{4) Runtime monitoring and continuous attestation.}  
Modules can opt into runtime verification hooks, providing consumption‑side attestation or telemetry. This model is informed by the transparency principles used in Sigstore’s Rekor log, where signatures are globally auditable and permanently recorded. Sigstore’s identity-based signing model is summarized in Fig.~\ref{fig:sigstore-flow}.

Together, these design elements enable AI-assisted development without sacrificing security or compositional trustworthiness, ensuring that generated systems remain within well-defined safety envelopes.
\section{Threat Model and Mitigation Strategies}

Human-Certified Module Repositories (HCMRs) are designed to counter a broad spectrum of threats that have repeatedly manifested across modern software supply chains. This section formalizes the threat model by identifying adversary capabilities, attack surfaces, and failure modes highlighted in recent empirical studies of real-world compromises.

\subsection{Adversary Capabilities}

\textbf{1) Compromise of build infrastructure.}  
The SolarWinds incident demonstrated that adversaries with access to CI/CD pipelines can inject malicious code into trusted update channels, propagating trojanized software to approximately 18{,}000 downstream organizations \cite{Willett2021SolarWindsSurvival,Martinez2021SolarWindsCaseStudy}. Reports show that attackers infiltrated the build system as early as 2019 and embedded malicious code into a signed Orion software update, leveraging the integrity of SolarWinds’ official distribution channel \cite{Coco2022SolarWindsEJIL,Ghanbari2024SolarWindsKaseyaTeachingCase}.

\textbf{2) Abuse of maintainership and contributor trust.}  
The 2024–2025 XZ Utils backdoor revealed that long-term social engineering can be used to infiltrate critical open-source projects. Analyses document how the attacker gradually gained maintainer-level trust before introducing a multi-stage backdoor targeting \texttt{sshd} through build-script manipulation and embedded payloads \cite{Lins2024XZArxiv,Kaspersky2024XZAnalysis}. Further research shows that the affected library was a dependency for nearly 30{,}000 Debian and Ubuntu packages, illustrating the severity of maintainer compromise \cite{Lins2025XZLNCS,MITSTAMP2025XZDeck}.

\textbf{3) Exploitation of widely deployed components.}  
Long-term studies of Log4Shell show sustained exploitation between 2021 and 2025, with attacker infrastructure evolving to include obfuscation, alternative protocols, and persistent scanning waves \cite{Singh2026Log4ShellLongitudinal}\cite{Hiesgen2024Log4jTNSM}. Attackers also reused infrastructure across years, indicating the feasibility of sustained, low-cost exploitation campaigns.

\textbf{4) Manipulation of provenance and signing signals.}  
Supply-chain attacks frequently leverage trusted signatures to mask malicious artifacts. For example, SolarWinds attackers delivered a malicious update that was signed using legitimate SolarWinds certificates, bypassing downstream verification mechanisms that relied solely on signature validity \cite{Newman2022SigstoreCCS,SigstoreOverviewDocs}. The SLSA framework highlights the danger of trusting unsigned or unverifiable build metadata and emphasizes the importance of strong, tamper-evident provenance \cite{SLSAProvenanceV01,SLSAProvenanceV02,SLSAGitHubGenerator}.

\textbf{5) Abuse of modular composition ecosystems.}  
Ecosystems such as IFTTT, which integrate hundreds of services, exhibit execution-path fragility and inefficiencies due to loosely governed triggers and actions, as measured in a large-scale empirical study \cite{Mi2017IFTTTIMC}. Flow-based systems such as Node-RED similarly demonstrate variability in quality and correctness of community-designed modules, increasing the attack surface for malicious or faulty component integration \cite{Clerissi2018NodeRedTesting,Onwuegbuzie2024NodeRedAnalytics,IEEE2018NodeREDHomeAutomation}.

\subsection{Threat Surfaces}

Drawing on these findings, HCMRs consider the following primary threat surfaces:

\begin{itemize}
    \item \textbf{Dependency Injection Attacks:} Attackers insert malicious libraries into dependency graphs, as documented in Log4Shell’s transitive exposure and XZ Utils’ integration into thousands of upstream packages \cite{Singh2026Log4ShellLongitudinal,Lins2025XZLNCS}.
    \item \textbf{Build-System Subversion:} Malicious actors tamper with build steps or inject payloads into build scripts, as demonstrated by the XZ backdoor’s multi-stage build pipeline manipulation \cite{Lins2024XZArxiv,Kaspersky2024XZAnalysis}.
    \item \textbf{Maintainer Account Hijacking:} Attackers gain contributor roles, influence code review pathways, or coerce maintainers, mirroring the techniques used in the XZ Utils compromise \cite{MITSTAMP2025XZDeck}.
    \item \textbf{Key or Credential Abuse:} Traditional signing approaches depend on long-lived private keys, which can be stolen or misused. Sigstore research highlights the security risks of manual key management and the benefits of ephemeral, identity-based signing \cite{Newman2022SigstoreCCS,Kalu2025SigstoreUsability}.
    \item \textbf{Insecure Automation:} Automated or AI-assisted workflows may integrate modules without verifying provenance, contract compatibility, or security posture—issues exemplified by behavioral inconsistencies observed in IFTTT workflow execution \cite{Mi2017IFTTTIMC}.
\end{itemize}

\subsection{Mitigation Summary}

HCMRs mitigate these risks by integrating:

\textbf{1) Strong provenance generation and verification.}  
SLSA-aligned provenance structures ensure that each module includes verifiable metadata describing builder identity, dependencies, and build parameters, preventing the type of opaque build manipulation seen in SolarWinds and XZ Utils \cite{SLSAProvenanceV01,SLSAProvenanceV02,SLSAGitHubGenerator}.

\textbf{2) Curated module review and certification.}  
Modules undergo human review for interface correctness, dependency hygiene, and potential abuse vectors. This approach parallels well-governed ecosystems such as Azure Verified Modules, which enforce strict quality, testing, and design controls across all released modules \cite{AVM_MicrosoftLearn,AVM_Portal,AVM_GitHubRepo,AVM_SolutionDev,AVM_MonthlyUpdateJan2024}.

\textbf{3) Identity-bound artifact signing.}  
By integrating Sigstore’s OIDC-based ephemeral signing model, HCMRs eliminate reliance on long-lived private keys and ensure that signing events are logged in transparency logs (e.g., Rekor) for global auditability \cite{Newman2022SigstoreCCS,SigstoreOverviewDocs}.

\textbf{4) Governance against maintainer compromise.}  
Learnings from the XZ Utils compromise motivate multi-party code review, mandatory provenance checks, and automated anomaly detection for suspicious contributor behavior \cite{MITSTAMP2025XZDeck}.

\textbf{5) Safe composition constraints for AI systems.}  
AI-assisted assembly pipelines must satisfy metadata constraints, preventing integration of unverified modules. This mitigates risks observed in ecosystems such as IFTTT and Node-RED where loosely governed modules cause execution inconsistencies \cite{Mi2017IFTTTIMC,Clerissi2018NodeRedTesting}.

The threat landscape relevant to HCMRs is summarized in Table~\ref{tab:threat-matrix}.

\begin{table}[!t]
\centering
\caption{Threat categories relevant to software supply-chain security and the corresponding HCMR mitigation mechanisms.}
\label{tab:threat-matrix}
\begin{tabular}{lc}
\toprule
\textbf{Threat Type} & \textbf{HCMR Mitigation} \\
\midrule
Dependency Injection Attacks     & Module Certification + Provenance \\
Build-System Subversion          & Verified Provenance \\
Maintainer Account Hijacking     & Multi-party Review \\
Key or Credential Abuse          & Ephemeral Signing \\
Insecure Automation              & Contract-Aware Composition \\
\bottomrule
\end{tabular}
\end{table}
\section{Case Studies and Motivating Incidents}

To illustrate how Human-Certified Module Repositories (HCMRs) address real-world failure modes, this section examines three representative supply-chain incidents: SolarWinds, Log4Shell, and the XZ Utils backdoor. Each case highlights a different breakdown in trust and demonstrates how HCMRs would mitigate similar future failures.

\subsection{Case Study 1: SolarWinds Orion Compromise}

Academic analyses show that attackers infiltrated the SolarWinds build environment in 2019, modifying the Orion platform’s source code to insert a covert backdoor that was later distributed through digitally signed updates \cite{Willett2021SolarWindsSurvival,Martinez2021SolarWindsCaseStudy}. The compromised update was downloaded by approximately 18{,}000 customers, including U.S. federal agencies and Fortune 500 firms, illustrating the massive blast radius of trusted update channels \textit{without provenance attestation} \cite{Coco2022SolarWindsEJIL}.

Post-incident reporting indicates that attackers exploited lax access controls and insufficient monitoring of build-system integrity, enabling malware insertion during CI/CD operations that appeared legitimate downstream because the resulting binaries were signed using authentic SolarWinds certificates \cite{Ghanbari2024SolarWindsKaseyaTeachingCase}.

\textbf{How HCMRs mitigate this failure:}  
HCMRs require SLSA-level provenance, ensuring that each artifact is accompanied by a verifiable build history, builder identity, and cryptographic digests of all dependencies \cite{SLSAProvenanceV01,SLSAProvenanceV02,SLSAGitHubGenerator}. This prevents the distribution of unsigned or opaque builds and ensures tampering is detectable. Additional constraints such as multi-party reviews and human certification of modules align with practices validated in curated module ecosystems like Azure Verified Modules \cite{AVM_MicrosoftLearn,AVM_Portal,AVM_GitHubRepo,AVM_SolutionDev,AVM_MonthlyUpdateJan2024}.

\subsection{Case Study 2: Log4Shell (CVE-2021-44228)}

The Log4Shell vulnerability allowed remote code execution in the ubiquitous Log4j library. A 2026 longitudinal measurement study using an active network telescope found that exploitation persisted for years after disclosure (2021–2025), with attackers reusing infrastructure, evolving payload obfuscation strategies, and shifting scanning behaviors based on global vantage points \cite{Singh2026Log4ShellLongitudinal}. A complementary IEEE Transactions study observed repeated waves of malicious scanning throughout 2022, further demonstrating that ecosystems remain vulnerable long after patches become available \cite{Hiesgen2024Log4jTNSM}.

The severity of Log4Shell stemmed from three factors: deep transitive dependency chains, inconsistent patch adoption across industries, and the lack of centralized governance over critical building-block libraries. These structural weaknesses enabled the vulnerability to remain exploitable long after vendor patches were issued.

\textbf{How HCMRs mitigate this failure:}  
HCMRs enforce the publication of dependency digests, interface contracts, and assurance levels. This allows automated systems—including AI assistants—to detect when modules depend on vulnerable versions of critical components. Provenance-aware build orchestration ensures that downstream modules cannot be certified unless they use patched and verified dependencies, preventing long-tail exposure.

\subsection{Case Study 3: XZ Utils Backdoor (CVE-2024-3094)}

In 2024, investigators uncovered a multi-stage backdoor in the XZ Utils compression library, introduced through socially engineered maintainer compromise. Analyses show that the attacker embedded malicious code within test files, modified build scripts to extract and link the payload, and targeted \texttt{sshd} to obtain unauthenticated remote code execution \cite{Lins2024XZArxiv}. Subsequent ecosystem-wide dependency analysis revealed that XZ Utils was used by nearly 30{,}000 Debian and Ubuntu packages, making this compromise one of the most severe supply-chain incidents in the history of Linux distributions \cite{Lins2025XZLNCS,MITSTAMP2025XZDeck}.

Kaspersky’s incident analysis reported that the backdoor was sophisticated, multi-staged, stealthy, and difficult to detect using traditional malware scanning techniques due to its integration into the build pipeline rather than source-code-level artifacts \cite{Kaspersky2024XZAnalysis}.

\textbf{How HCMRs mitigate this failure:}  
HCMRs eliminate single-maintainer trust by enforcing multi-party review, provenance verification, and automated monitoring of contributor behavior. Modules undergo sandboxed behavioral validation, preventing malicious build-time extraction or linking logic from being certified. Transparency logs (as used in Sigstore) ensure that all signing actions are publicly verifiable, making covert backdoor activation significantly more difficult \cite{Newman2022SigstoreCCS,SigstoreOverviewDocs,Kalu2025SigstoreUsability}.

\subsection{Summary}

These case studies reveal complementary facets of the modern supply-chain threat landscape:  
(1) build-system tampering,  
(2) transitive dependency risk, and  
(3) maintainer compromise.  

HCMRs provide strong, systematic mitigations by combining provenance attestation, module certification, secure signing, curated governance, and AI-aware composition constraints—all informed by empirical research across SolarWinds, Log4Shell, XZ Utils, SLSA, Sigstore, and formal verification ecosystems \cite{Klein2014seL4TOCS,Leroy2009CompCertCACM}.
\section{Discussion and Implications}

The introduction of Human-Certified Module Repositories (HCMRs) reflects a deeper structural shift in software engineering. Historically, correctness guarantees and security assurances were achieved through extensive testing, static analysis, or formal verification. Systems such as the seL4 microkernel demonstrated that machine-checked proofs can secure critical kernel components end-to-end, maintaining verification even as the code evolved over a decade \cite{Klein2014seL4TOCS}. Verified compilation, exemplified by CompCert, similarly eliminates miscompilation bugs that could undermine software correctness at the binary level \cite{Leroy2009CompCertCACM}. Yet these techniques remain too expensive to scale across the long‑tail of software modules that modern systems depend on.

At the same time, high-impact supply-chain incidents have demonstrated that the primary risk in modern software ecosystems lies not in individual codebases but in \emph{the relationships between them} \cite{Coco2022SolarWindsEJIL,Ghanbari2024SolarWindsKaseyaTeachingCase}. The SolarWinds compromise showed that infiltrating a single vendor’s build infrastructure can insert trojanized code into signed updates distributed to 18{,}000 organizations via trusted channels \cite{Willett2021SolarWindsSurvival,Martinez2021SolarWindsCaseStudy}. Similarly, long-term measurements of Log4Shell exploitation revealed sustained attacker interest over multiple years, including infrastructure reuse and evolving payload obfuscation techniques \cite{Singh2026Log4ShellLongitudinal}. The XZ Utils compromise introduced a novel attack vector: social engineering of maintainers followed by multi-stage build-time backdoor insertion, ultimately affecting nearly 30{,}000 downstream packages in major Linux distributions \cite{Lins2024XZArxiv,Lins2025XZLNCS,MITSTAMP2025XZDeck,Kaspersky2024XZAnalysis}.

Traditional package repositories (e.g., npm, PyPI, Maven Central) are optimized for openness and scale. They rely on the good faith of maintainers and downstream users, lacking structural safeguards against sophisticated supply‑chain attackers. In contrast, curated ecosystems such as Azure Verified Modules enforce design standards, code reviews, automated testing, versioning discipline, and architectural compliance, ensuring consistent quality across module releases \cite{AVM_MicrosoftLearn,AVM_Portal,AVM_GitHubRepo,AVM_SolutionDev,AVM_MonthlyUpdateJan2024}. However, AVM focuses on cloud infrastructure modules rather than general‑purpose software components, and does not include explicit mechanisms for handling social‑engineering threats, provenance tampering, or AI‑based composition.

HCMRs seek to bridge this gap by providing a scalable, security-centric, provenance‑aware foundation for modular software construction. Their design is informed by lessons from provenance frameworks such as SLSA, which define structured, tamper‑evident metadata capturing builder identity, build configuration, and dependency digests \cite{SLSAProvenanceV01,SLSAProvenanceV02}. Similarly, the Sigstore ecosystem demonstrated that identity‑based signing with ephemeral keys and transparency logs can reduce key‑management overhead while improving auditability of software artifacts \cite{Newman2022SigstoreCCS,SigstoreOverviewDocs,Kalu2025SigstoreUsability}.

The rise of AI‑assisted development introduces new considerations. Tools like GitHub Copilot and similar LLM‑powered agents are capable of assembling multi‑component systems rapidly, but may inadvertently select vulnerable or unverified dependencies if left unconstrained. Studies of IFTTT’s trigger–action ecosystem reveal how even simple composition models can exhibit performance inconsistencies and unintended interactions when modules are loosely governed \cite{Mi2017IFTTTIMC}. Node‑RED research similarly shows that visual flow composition increases flexibility but creates validation challenges that require structured testing approaches \cite{Clerissi2018NodeRedTesting,Onwuegbuzie2024NodeRedAnalytics,IEEE2018NodeREDHomeAutomation}.

These findings highlight the importance of metadata‑driven, contract‑aware assembly constraints—centralized in the HCMR design.

Ultimately, HCMRs represent a synthesis of multiple research trajectories: formal verification for correctness, provenance frameworks for integrity, curated module ecosystems for consistency, and supply‑chain security literature for threat modeling. Their goal is not to replace existing repositories but to add a security‑assured layer for modules intended for critical systems or AI‑driven composition.

\subsection{Typed Languages as Compiler-Enforced Guardrails for AI-Generated Code}
\label{subsec:typed-guardrails}

As AI assistants generate a growing share of application code, correctness and security defects increasingly arise from structural inconsistencies—mismatched data contracts, silent type coercions, aliasing and lifetime errors, or missing validation logic—rather than from simple syntax errors. Strongly typed languages and strict compilers provide an architectural countermeasure: they force AI-generated code to satisfy type and interface constraints before execution, turning compilation into a first-class safety filter. Recent industry analysis argues that as “code you did not personally write” increases, static type systems act as an essential safety net that surfaces ambiguous logic early; GitHub reports that the majority of LLM-generated compilation failures are type-check errors and that typed ecosystems reduce downstream fixes in AI-driven workflows \cite{GitHubBlog2026Typed}. Independent assessments similarly emphasize that typed languages reduce ambiguity, making code easier for both AI tools and humans to reason about by providing explicit intent and predictable data structures \cite{TechBusinessNews2026TypedShift}.

\paragraph{Compiler-enforced structure against silent failure}
Dynamic languages allow rapid prototyping, but their permissiveness lets AI-generated code ``look plausible, run immediately, and fail silently.'' Typed ecosystems mitigate this failure mode: a type error blocks execution until contracts are reconciled. Developer discourse around “vibe coding’’ frequently identifies typed languages (e.g., TypeScript, Go, Rust) as more robust when co-generating code with LLMs, precisely because they convert latent runtime failures into actionable compile-time diagnostics \cite{SignalsVibeDebate2025}. Rust is often highlighted as a paradigmatic example: its ownership and lifetime model enforces memory-safety invariants that an LLM must satisfy before code is accepted. As a result, compilation failures become iterative guidance, and a successfully compiling Rust program is more likely to be correct \cite{Forbes2026RustVibeCoding}.

\paragraph{Empirical security motivation}
Security studies consistently find that AI-generated code contains impactful flaws even when superficially functional. Industry analyses reveal widespread omission of validation, missing access controls, and unsafe patterns that reflect training-data weaknesses \cite{CSA2025AIGeneratedRisks}. A policy study from CSET similarly reports that nearly half of evaluated LLM outputs contain bugs with security implications, arguing for structured tool-assisted validation \cite{CSET2024AICodeRisks}. Furthermore, controlled experiments show that iterative “refinement’’ with LLMs can worsen security: vulnerability prevalence increases during multi-round improvement cycles unless external constraints are applied \cite{Shukla2025SecurityDegradation}. Typed languages complement static analysis and human review by eliminating entire classes of errors by construction.

\paragraph{Alignment with HCMRs}
Typed languages align naturally with HCMR contract and provenance metadata. Interface contracts can map directly onto language-level types or specification stubs, while compilers provide a conformance oracle that AI agents must satisfy prior to module integration. In effect, typing operationalizes HCMR guarantees at build time: provenance ensures artifacts are traceable, certification ensures modules are trustworthy, and strong typing ensures that AI-produced glue code is structurally sound. Together, these mechanisms help constrain AI-driven composition to safe, predictable, and verifiably correct integration paths.

\subsection{Industry Responses and Emerging Security Architectures}

Recent developments in AI-assisted software engineering indicate that the need for security-aware composition is increasingly recognized beyond academic research. Commercial and open-source tooling has begun to embed security intelligence directly into AI coding workflows, reflecting a shift toward integrating security checks at the moment code is generated or assembled. A prominent example is Endor Labs' AURI platform, which introduces a lightweight security-intelligence layer into IDEs and agentic coding assistants, aiming to detect vulnerabilities and exposed secrets as code is produced rather than after deployment \cite{VentureBeatAURI,EndorAURI}.

These industry directions echo empirical findings in the research community regarding the gap between functional correctness and security in AI-generated artifacts. A recent multi-institution study led by Carnegie Mellon University introduces the SUSVIBES benchmark to evaluate AI coding agents under realistic constraints. The authors report that while leading agents achieve functional correctness on roughly 61\% of tasks, only 10.5\% of these outputs satisfy both functional and security requirements, leaving the majority of AI-generated patches still vulnerable \cite{SUSVIBES2025}. This result highlights a structural shortfall that stems not merely from code-level patterns but from missing context about dependency relationships, reachable attack surfaces, and the security properties of reused components.

Taken together, these industry and academic developments underscore a convergent trend: as AI systems assume a greater role in code generation and software assembly, security must be embedded directly into the composition process. Rather than treating security as a downstream validation step, emerging tools and studies point toward architectures that provide security constraints, provenance, and interface guarantees at generation time. Human-Certified Module Repositories (HCMRs) align with this trajectory by furnishing a curated ecosystem of modules whose provenance, behavioral contracts, and security review can be leveraged both by humans and by AI coding agents to reduce the likelihood of unsafe compositions.
\section{Open Research Challenges}

While HCMRs present a promising direction for securing modular, AI‑constructed systems, several foundational challenges remain open for research.

\subsection{Scalable Certification}

The XZ Utils compromise demonstrated that even widely deployed components may be maintained by a single overburdened volunteer, creating a systemic point of failure \cite{Lins2024XZArxiv,Lins2025XZLNCS,MITSTAMP2025XZDeck}. Scaling human certification to thousands of modules requires advances in automated reasoning, static analysis, formal specification mining, and ML‑assisted review workflows. Determining where human oversight is essential versus where automated checks suffice remains an open question.

\subsection{Formal Contracts for Composability}

Empirical analyses of ecosystems such as IFTTT show that loosely defined trigger–action interfaces frequently lead to unpredictable or inefficient runtime behavior due to unclear semantics and inconsistent execution guarantees \cite{Mi2017IFTTTIMC}. Likewise, studies of Node‑RED highlight the difficulty of validating correctness in visually composed flows without explicit formal models \cite{Clerissi2018NodeRedTesting,Onwuegbuzie2024NodeRedAnalytics,IEEE2018NodeREDHomeAutomation}. Research is needed to design expressive, machine‑checkable interface contracts that support both human readability and automated enforcement.

\subsection{Detecting Maintainer Compromise}

The multi‑year social‑engineering campaign behind the XZ Utils backdoor illustrates the need for contributor‑behavior anomaly detection, trust‑score modeling, and multi‑party approval systems that reduce reliance on single maintainers \cite{Lins2024XZArxiv,Kaspersky2024XZAnalysis,MITSTAMP2025XZDeck}. Techniques such as reputation systems, code‑review graph analysis, and automated semantic differencing could help detect suspicious behavioral patterns early.

\subsection{Provenance-Enforced Build Reproducibility}

While SLSA defines rigorous provenance structures, widespread adoption remains a challenge. Many ecosystems lack automated tooling to generate or verify in-toto attestations at scale. Research is needed to integrate reproducible builds, cryptographic dependency digests, and incremental verification workflows into mainstream development pipelines, enabling assurance levels equivalent to those envisioned in the SLSA framework \cite{SLSAProvenanceV01,SLSAProvenanceV02,SLSAGitHubGenerator}.

\subsection{AI Accountability and Guardrails}

As AI-assisted code generation becomes standard, HCMRs must provide mechanisms for ensuring that AI systems select certified modules, adhere to contract constraints, and generate verifiable provenance. Studies of usability barriers in software-signing ecosystems such as Sigstore indicate that adoption hinges on developer-friendly tooling and integration pathways \cite{Newman2022SigstoreCCS,SigstoreOverviewDocs,Kalu2025SigstoreUsability}. Similar work is required to design AI‑facing guardrails that prevent unsafe module compositions.

\subsection{Governance and Evolution}

Curated ecosystems such as Azure Verified Modules show the value of centralized quality standards, versioning discipline, and architectural compliance \cite{AVM_MicrosoftLearn,AVM_Portal,AVM_GitHubRepo,AVM_SolutionDev,AVM_MonthlyUpdateJan2024}. However, governance models for HCMRs require additional mechanisms for transparent dispute resolution, deprecation, cross‑module dependency policy, and long-term maintainability. Research is needed to define sustainable governance frameworks that can evolve with ecosystem growth.

These open problems highlight the need for interdisciplinary work that bridges security, formal methods, programming languages, distributed systems, and human–computer interaction to fully realize the potential of HCMRs.
\section{Conclusion}

The software ecosystem is undergoing a fundamental transformation driven by increasingly complex dependency graphs, rapidly evolving threat actors, and the growing influence of AI-assisted development. Recent supply-chain incidents such as SolarWinds, Log4Shell, and the XZ Utils backdoor reveal the fragility of current trust models, demonstrating how attackers can exploit build pipelines, maintainer accounts, or ubiquitous open-source libraries to achieve wide-ranging compromise. SolarWinds alone propagated a malicious update to approximately 18{,}000 organizations through a signed distribution channel, highlighting the systemic risk of opaque build processes \cite{Willett2021SolarWindsSurvival,Martinez2021SolarWindsCaseStudy,Coco2022SolarWindsEJIL,Ghanbari2024SolarWindsKaseyaTeachingCase}. Long-term measurement studies show that Log4Shell remained under active exploitation for years after disclosure, underscoring the persistence of ecosystem-level vulnerabilities \cite{Singh2026Log4ShellLongitudinal,Hiesgen2024Log4jTNSM}. The XZ Utils incident further revealed how attacker infiltration of maintainer trust can lead to backdoor implantation across tens of thousands of dependent packages \cite{Lins2024XZArxiv,Lins2025XZLNCS,Kaspersky2024XZAnalysis,MITSTAMP2025XZDeck}.

In parallel, curated ecosystems such as Azure Verified Modules offer evidence that structured governance, testing discipline, and architectural consistency can improve reliability and user trust at scale \cite{AVM_MicrosoftLearn,AVM_Portal,AVM_GitHubRepo,AVM_SolutionDev,AVM_MonthlyUpdateJan2024}. Provenance frameworks such as SLSA provide structured, tamper-evident metadata linking artifacts to their build systems, enabling systematic verification of software integrity \cite{SLSAProvenanceV01,SLSAProvenanceV02,SLSAGitHubGenerator}. Sigstore's identity-based signing model further lowers the barrier to authenticated, transparent artifact signing, mitigating key-management burdens and improving auditability \cite{Newman2022SigstoreCCS,SigstoreOverviewDocs,Kalu2025SigstoreUsability}.

Human-Certified Module Repositories synthesize these advances into a new secure substrate for modular software development. By combining human certification, provenance verification, contract-enforced composition, and secure-by-default assembly pipelines, HCMRs provide a foundation for both human and AI agents to build systems from trustworthy components. They enable predictable, accountable, and auditable software construction in a world where automation and interdependence are the norm.

HCMRs are not a replacement for existing package repositories; instead, they form a complementary high-assurance layer designed for critical applications, safety-sensitive domains, and AI-driven code synthesis. They address the unique challenges posed by modern supply-chain threats and prepare the ecosystem for an era where developers—both human and artificial—must assemble systems from components with clearly established trust, provenance, and guarantees.

As software ecosystems continue to scale, the principles underlying HCMRs—curation, transparency, provenance, and compositional assurance—will be essential in enabling secure, reliable, and predictable software for the next generation of systems.

\phantomsection
\pdfbookmark[1]{Acknowledgment}{acknowledgment}
\section*{Acknowledgment}
The author acknowledges the use of GPT-5.3 to assist with grammar correction,
wording refinement, and readability improvements throughout the manuscript.
The author reviewed and edited the AI-assisted text and assumes full
responsibility for the final content of the article.

\balance
\bibliographystyle{IEEEtran}
\bibliography{references}

\end{document}